\documentclass[preprint]{aastex}
\usepackage{graphicx}
\usepackage{color}
\usepackage{natbib}
\usepackage{amsmath}
\usepackage[nomarkers]{endfloat}

\shorttitle{Color All-Sky Panorama Image}
\shortauthors{A. Mellinger}

\begin{document}


\title{A Color All-Sky Panorama Image of the Milky Way}


\author{Axel Mellinger}
\affil{Department of Physics, Central Michigan University, Mount Pleasant, MI 48859}






\begin{abstract}
This article describes the assembly of an optical (RGB) all-sky mosaic image with an image scale of 36~arcsec/pixel, a limiting magnitude of approx.\ 14~mag and an 18~bit dynamic range. Using a portable low-cost system, 70~fields (each covering $40\degr\times 27\degr$) were imaged over a time span of 21 months from dark-sky locations in South Africa, Texas and Michigan. The fields were photometrically calibrated against standard catalog stars.  Using sky background data from the Pioneer 10 and 11 space probes, gradients resulting from artificial light pollution, airglow and zodiacal light were eliminated, while the large-scale galactic and extragalactic background resulting from unresolved sources was preserved. The 648~Megapixel image is a valuable educational tool, being able to fully utilize the resolution and dynamic range of modern full-dome planetarium projection systems. 
\end{abstract}





\section{Introduction}


Since ancient times, astronomers have sought to draw maps of the sky, similar to their cartographer colleagues' drive to produce ever more accurate maps of the Earth. 
However, it was not until the invention of photography that faint, large-scale structures of nebulae and dust lanes could be faithfully reproduced. Since the late 19th century, several visible-light, wide-angle photographic surveys of the sky have been completed, such as E.~E.~Barnard's \textit{Atlas of Selected Regions of the Milky Way}~\citep{Bar1927}, the National Geographic Society-Palomar Observatory Sky Survey of the northern sky~\citep{Min1963} and subsequently the equivalent ESO/SERC atlas of the southern sky, and more recently the Sloan Digital Sky Survey \citep{York2000}. In addition, a H$\alpha$ narrow-band optical survey was compiled~\citep{Fin2003}.

In 2000, the author published an all-sky color image taken on photographic 35~mm color film with a focal length of $f=28$~mm. A total of 51 images were computer-processed into a mosaic measuring $14\,400\times 7\,200$ pixels, equivalent to an image scale of 90~arcsec/pixel \citep{diCicco1999,Kiz2001}. This picture is now widely used for educational purposes in planetariums, exhibitions and textbooks, formed the basis for a photographic star atlas \citep{Mel2005} and illustrates the location of star-forming regions in the Milky Way \citep{Mel2008}.
With recent advances in digital projection technology, a demand grew for both a higher spatial resolution and a higher dynamic range. The article describes the assembly of a 648 MPixel mosaic from more than 3000 individual CCD images.

\section{Data Acquisition}
\subsection{Hardware}
\subsubsection{CCD camera}
Wide-field imaging equipment based on commercial camera lenses has been successfully used in the past for, e.\,g., emission-line surveys~\citep{SHASSA} and measurements of the night-sky brightness~\citep{Sha2005,Dur2007}.
After exploring several camera systems on the market, an SBIG STL-11000 CCD camera was chosen due to its
\begin{itemize}
\item moderate cost,
\item large sensor size (36~mm $\times$ 24~mm), 
\item anti-blooming gate (eliminating the need for software de-blooming), and
\item compact construction (low back focus, allowing the use of 35~mm camera lenses).
\end{itemize}
The camera uses a Kodak KAI-11002 chip, a 4008 (H) $\times$ 2672 (V) progressive scan interline CCD image sensor with an anti-blooming gate~\citep{Kodak}. It uses micro-lenses to enhance its quantum efficiency by directing light to the active pixel areas. Drawbacks of this design include non-linearity and angular quantum-efficiency variations, and are addressed in sections \ref{sec:nonlin} and \ref{sec:preproc}.

\subsubsection{Lens}
The camera was fitted with a Minolta MD 1.4/50~mm lens originally used on a 35~mm format film-based single-lens reflex camera. To improve the image quality, the lens was stopped down to $f/4$ for all exposures.

\begin{figure}
\includegraphics[width=\columnwidth]{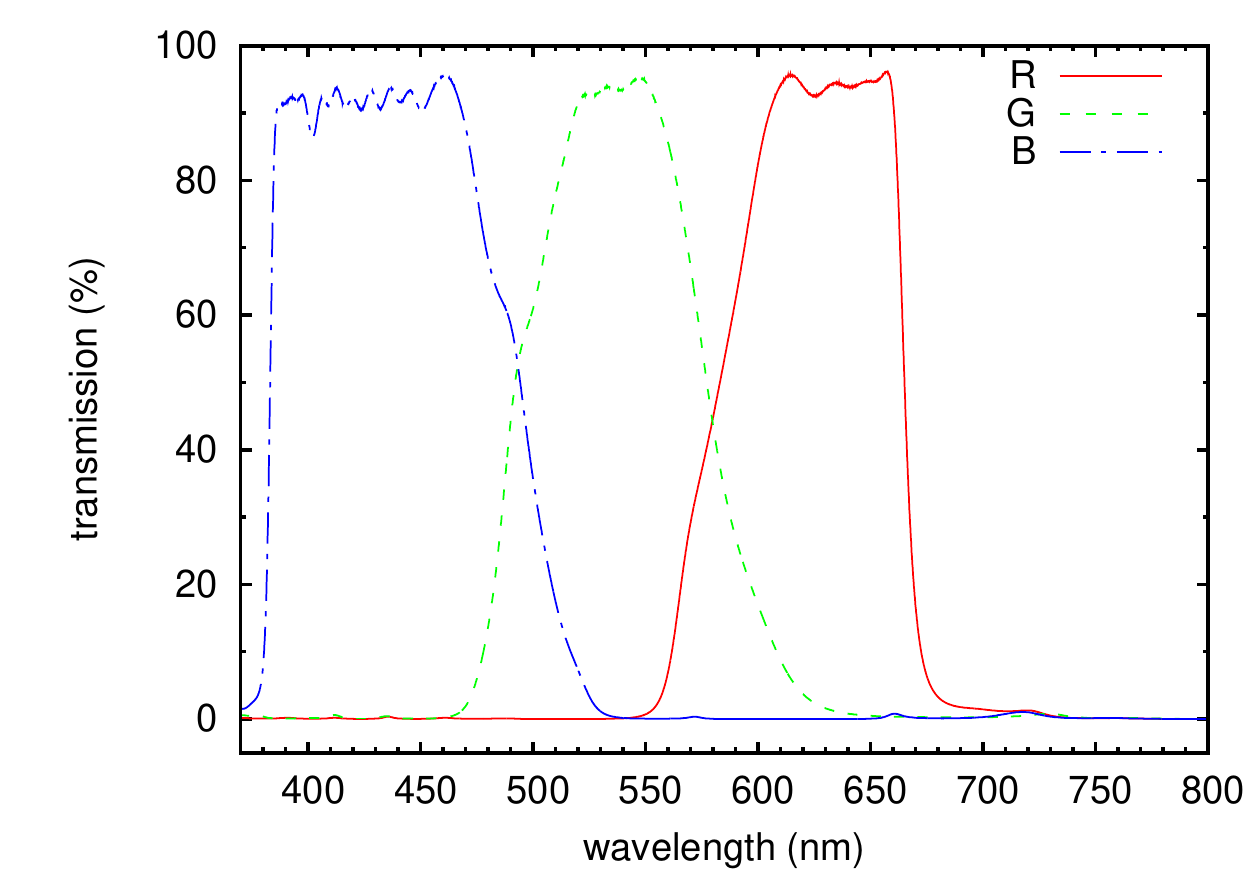}
\caption{Transmission curves of the Astronomik RGB filters.}
\label{fig:RGBcurves}
\end{figure}
\subsubsection{Filters}
As the principal purpose of the panorama image is to create a visual representation of colors in the sky, an RGB filter set (Astronomik Type 2c) was used instead of the Johnson UBVR filters common in photometry. The transmission curves are shown in Fig.~\ref{fig:RGBcurves}.

\subsection{Field Selection}
\begin{figure*}
\includegraphics[width=\textwidth]{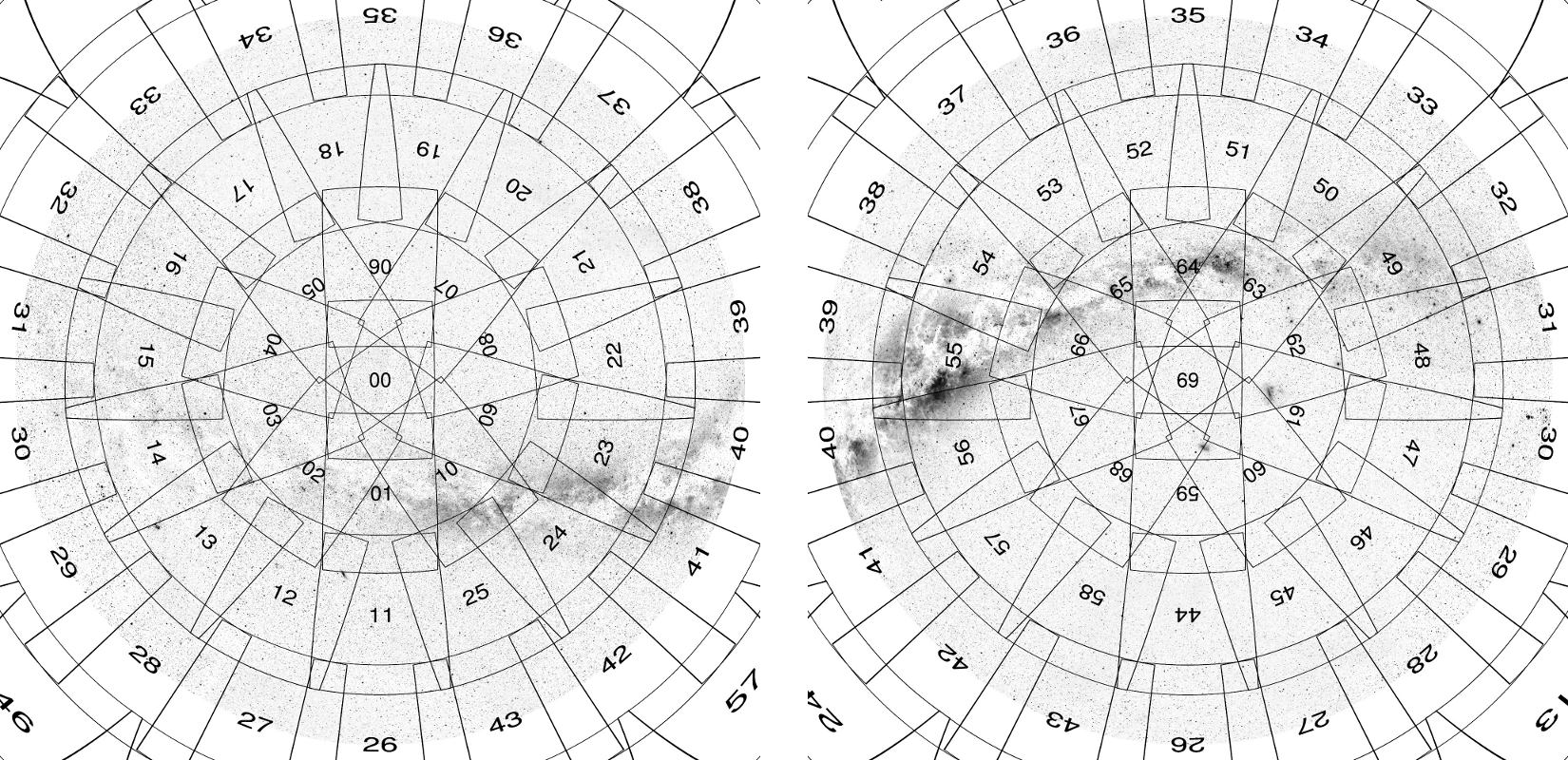}
\caption{Northern (left) and southern (right) hemisphere coverage maps of the 70 fields.}
\label{fig:fieldmap}
\end{figure*}

The KAI-11002 chip has an active area of $36\times 24$~mm$^2$, resulting in a field of view of approx.\ $40\degr \times 27\degr$ at a focal length of $f=50$~mm. Due to the location of the mounting thread on the camera, a ``portrait'' orientation of the camera was chosen, with the long side of the chip being parallel to circles of constant right ascension. To ensure a seamless mosaic, adjacent fields had to overlap each other by at least 15\%\ of the respective offset in right ascension or declination. With these constraints, a coverage pattern (Fig.~\ref{fig:fieldmap}) was developed by way of trial-and-error. The pattern consists of five
rings of frames centered at declinations of $+62\degr$, $+32\degr$, $0\degr$, $-32\degr$ and $-62\degr$, in addition to two fields centered at the celestial poles ($\pm 90\degr$). The number of fields and offsets in right ascension are listed in Table~\ref{tab:fields}.

\begin{deluxetable}{crc}
\tablecaption{Declination of field centers and RA offsets\label{tab:fields}}
\tablehead{\colhead{Field number(s)} & \colhead{$\delta_\text{c}$} & \colhead{RA offset}}
\tablewidth{0pt}
\startdata
0     & $+90\degr$ & --\\
 1-10 & $+62\degr$ & $2^\text{h} 24^\text{m}$ \\
11-25 & $+32\degr$ & $1^\text{h} 36^\text{m}$ \\
26-43 &   $0\degr$ & $1^\text{h} 20^\text{m}$ \\
44-58 & $-32\degr$ & $1^\text{h} 36^\text{m}$ \\
59-68 & $-62\degr$ & $2^\text{h} 24^\text{m}$ \\
69    & $-90\degr$ & --
\enddata
\end{deluxetable}

%

\subsection{Exposure times}
In order to increase the dynamic range beyond the 16~bits of the camera's analog-to-digital converter (of which approx. 12~bits provide data above the noise level), three different exposure times (240~s, 15~s and 0.5~s) were used. Five frames were taken for each exposure time and filter setting.

\subsection{Sites}
The fields centered at declinations $-90\degr$, $-62\degr$, $-32\degr$ and $0\degr$ were taken in October 2007 and March/April 2008 from dark-sky locations in South Africa (Tab.~\ref{tab:sites}), while the northern hemisphere images (declinations $+32\degr$, $+62\degr$ and $+90\degr$) were taken between December 2008 and August 2009 from Texas and Michigan in the U.S.

\section{Data Processing}
\subsection{Non-linearity correction}\label{sec:nonlin}
\begin{figure}
\includegraphics[width=\columnwidth]{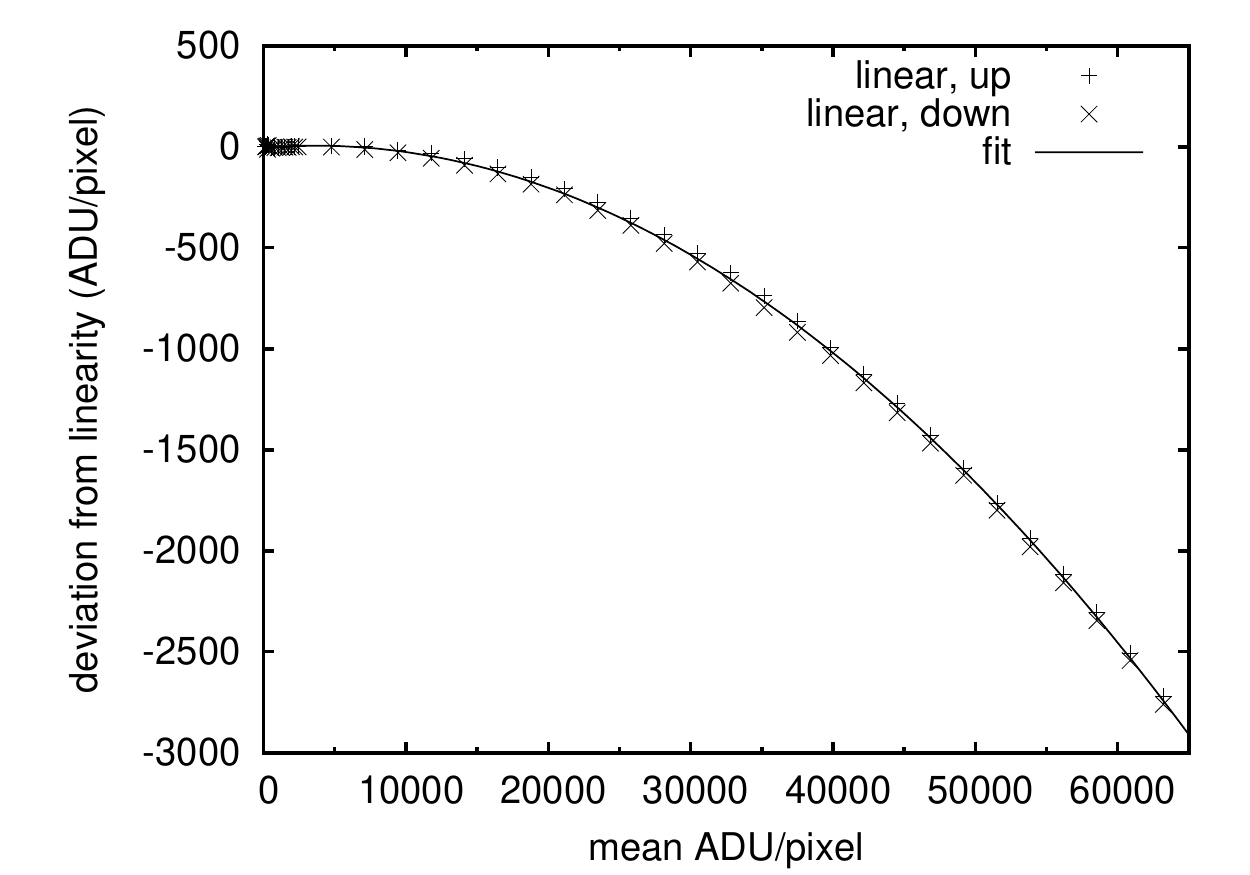}
\caption{Non-linearity of the STL-11000 camera obtained in laboratory tests with a constant-current LED light source. The solid line shows the least squares fit according to Eq.~(\ref{eq:model}).}
\label{fig:nonlin}
\end{figure}
The use of a CCD sensor with an anti-blooming gate often comes at the expense of linearity~\citep{Dur2007}.
To test the sensor linearity, a uniform white target illuminated by a light-emitting diode (LED) operated in constant-current mode was photographed with a series of exposure times ranging from 0.02 to 30~s. To check for variations in the LED intensity, two series of exposures (one from short to long exposure times, the other from long to short) was taken. After dark-frame subtraction, the flux in a field measuring $200\times 200$ pixels at the image center was measured using the IRAF \texttt{imstat} task. The test revealed a significant deviation from a linear response for a signal as low as 15000 ADUs pixel$^{-1}$, well below the full-well depth of approx.\ 60000 ADUs pixel$^{-1}$~(Fig.\ \ref{fig:nonlin}). However, the observed signal as a function of illumination could be well represented by the quadratic equation
\begin{equation}
 S(E) = a_1 E + a_2 E^2\:,
\label{eq:model}
\end{equation}
where $E$ is the exposure (product of light intensity and exposure time) and $S$ is the measured signal. 
Based on this data, a non-linearity correction was developed as follows:
Solving Eq.~(\ref{eq:model}) for $E$ yields
\[ E = \frac{-a_1 + \sqrt{a_1^2+4a_2S}}{2a_2} \]
The corrected (linearized) signal is then
\begin{eqnarray}
S' &=& a_1 E \nonumber\\
   &=& \frac{1}{2\alpha} \left( \sqrt{1+4\alpha S} - 1\right) \:,
\end{eqnarray}
where
\begin{equation}
\alpha := \frac{a_2}{a_1^2}
\end{equation}
From a least squares fit, $\alpha$ was determined as $-7.5334\times 10^{-7}$~ADU$^{-1}$. This function was applied to all raw images, including the flat frames (but not the dark frames, due to their low ADU values).

\subsection{Pre-processing}\label{sec:preproc}
\begin{figure}
\includegraphics[width=\columnwidth]{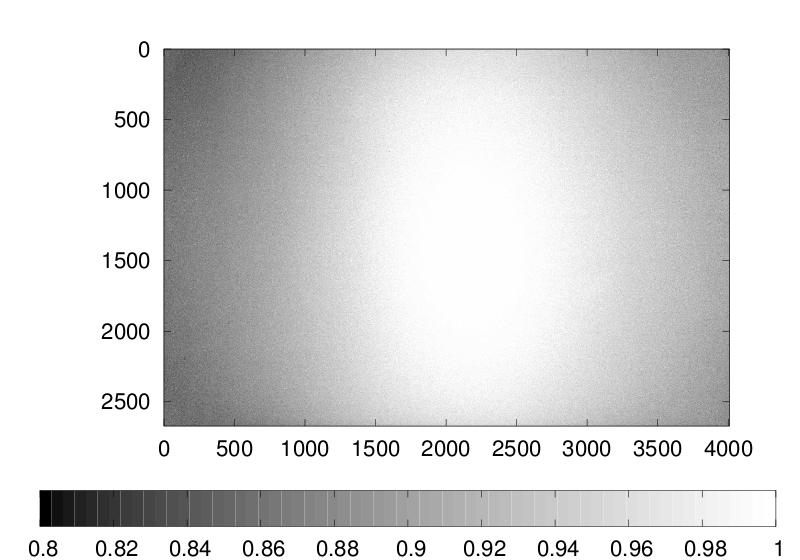}
\caption{Ratio of R and G flat fields.}
\label{fig:flats}
\end{figure}

The standard pre-processing steps (dark-frame subtraction, masking of bad pixels, division by flat frame) were applied to all images. Due to its micro-lenses, the angular quantum efficiency of the CCD chip depends upon the angle of incidence. This angular profile in turn is a function of the wavelength \citep{Kodak}. Hence, flat frames taken through the R, G and B filters differ significantly (Fig.~\ref{fig:flats}).
Depending on the spectral distribution of the sky background, this wavelength-dependent sensitivity of the camera system may lead to residual gradients, even with proper flat-frame calibration. However, these gradients were later eliminated by the background-subtraction procedure detailed in section~\ref{sec:bgsubt}.

Next, the images were upsampled by a factor of 3. Although it increases the file size by $9\times$, this step was found to be necessary in order to minimize the loss of resolution in subsequent resampling with bilinear interpolation. Higher order interpolation methods, such as bicubic or Lanczos resampling could not be used as they  introduce ringing artifacts in undersampled images.

\subsection{Astrometric Calibration}
Wide-angle imaging requires careful astrometric calibration, taking into account distortions of the camera lens beyond a simple gnomonic projection. 
In each raw image, up to 1500 reference stars were automatically selected using the Source Extractor software~\citep{Ber1996} and pattern-matched against sources in the Hubble Guide Star Catalog~1.2~\citep{Mor2001}. The pattern-matching algorithm was invented by \citet{Val1995} and implemented in C by \citet{Ric2008}. With these reference stars, an astrometric solution was calculated using custom software, and a world coordinate system (WCS) header was added to the FITS file. For each field, all images with a common exposure time and filter setting were then resampled and co-added with the SWarp software~\citep{Ber2002}. Finally, all pixels in the 240~s frames that were masked as saturated (cf.\ section~\ref{sec:preproc}) were replaced by scaled pixel values from the 15~s or 0.5~s frames using the IRAF \texttt{imexpr} task.

\subsection{Photometric Calibration}
\begin{figure*}
{\centering\includegraphics[width=\textwidth]{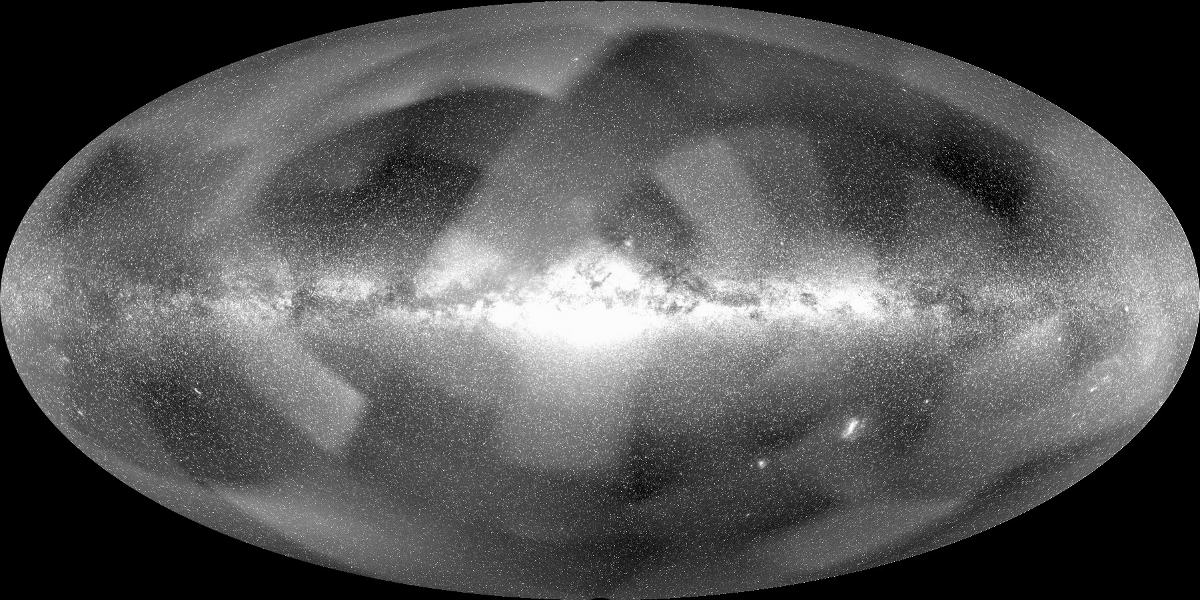}\\}
\caption{R channel mosaic, combined without photometric calibration and background subtraction. There are strong variations in the background level, as well as intra-field gradients.}
\label{fig:mosaic_nobg_noeq}
\end{figure*}

A major problem was the photometric calibration of the individual fields. A straight combination without photometric corrections yields a patchy-looking mosaic, as shown in Fig.~\ref{fig:mosaic_nobg_noeq}. Clearly, the sky background level varies significantly between fields, and some fields exhibit strong intra-field background gradients. Moreover, differences in atmospheric transparency may cause the recorded flux $I$ of a star to vary between two frames. 

\subsubsection{Contributions to the sky background}
The diffuse night-sky background has a variety of sources (see \citet{Lei1998} for an in-depth discussion):
\begin{enumerate}
 \item \textit{Light pollution ($I_\text{LP}$).} Although most images were taken from dark-sky sites (class 1 on Bortle's scale \citep{Bor2001}), time constraints dictated that some fields were taken from sites with slight artificial light pollution (the Michigan sites have Bortle class 2 or 3). In addition, again due to time constraints, some images were taken in the evening sky before the end of astronomical twilight, or in the morning sky past the beginning of astronomical twilight.

\item \textit{Airglow ($I_\text{A}$).} In the absence of light pollution, airglow is the largest contributor to the night sky brightness \citep{Lei1998}. 
In the visible spectrum, the main sources are atmospheric O$_2$ (for wavelengths $\lambda<430$~nm) and OH (for $\lambda>520$~nm). Air glow is highly variable both in spatial distribution and in time.

\item \textit{Zodiacal light ($I_\text{ZL}$).}  Zodiacal light is sunlight scattered by interplanetary dust particles. Near the ecliptic poles its intensity in the visible spectrum is typically 2-4 times weaker than airglow, but gains significantly as the field of view gets closer to the sun.
Due to its spatial confinement to the ecliptic region, it can cause strong background gradients in wide-field images.

\item \textit{Tropospheric scattering ($I_\text{sca}$).} Starlight, as well as airglow and zodical are scattered in the troposphere. High clouds, as well as high humidity at ground level can enhance scattering and lead to halos around bright stars.

\item \textit{Integrated starlight ($I_\text{ISL}$).} Another contribution to the sky brightness is the light from from unresolved stars. At visible wavelengths, it is dominated by main sequence stars~\citep{Lei1998}.

\item \textit{Diffuse galactic light ($I_\text{DGL}$).} Diffuse Galactic Light (DGL) is produced when starlight is scattered by dust grains in interstellar space.

\item \textit{Extragalactic background light ($I_\text{EBL}$)}. Light from galaxies that are not individually detected.
\end{enumerate}

Many of these contributions to the night sky brightness are undesirable in an all-sky mosaic image. Light pollution, airglow, zodiacal light and tropospheric scattering vary between individual fields, as well as within a field, and thus should be removed in the calibration process. However, the contribution from unresolved stars and the diffuse galactic light are features of the Milky Way and should be preserved, together with any extragalactic background. This creates a major calibration problem, as we need to decide on how much of the local background is caused by unwanted sources (1-4) or galactic and extragalactic contributions to be retained (5-7). This problem has plagued a number of ground-based studies of the galactic background~\citep{Lei1998}. However, these effects can be separated using space-borne visual photometry data from the Imaging Photopolarimeters (IPPs) onboard the Pioneer 10 and 11 space probes. Their data was obviously not contaminated by terrestrial sources, and the zodiacal light contribution became negligible for distances above 3.3~AU~\citep{Gor1997, Gor1998}.

The Photometric calibration of the all-sky image was performed in two steps:

\begin{figure}
\includegraphics[width=\textwidth]{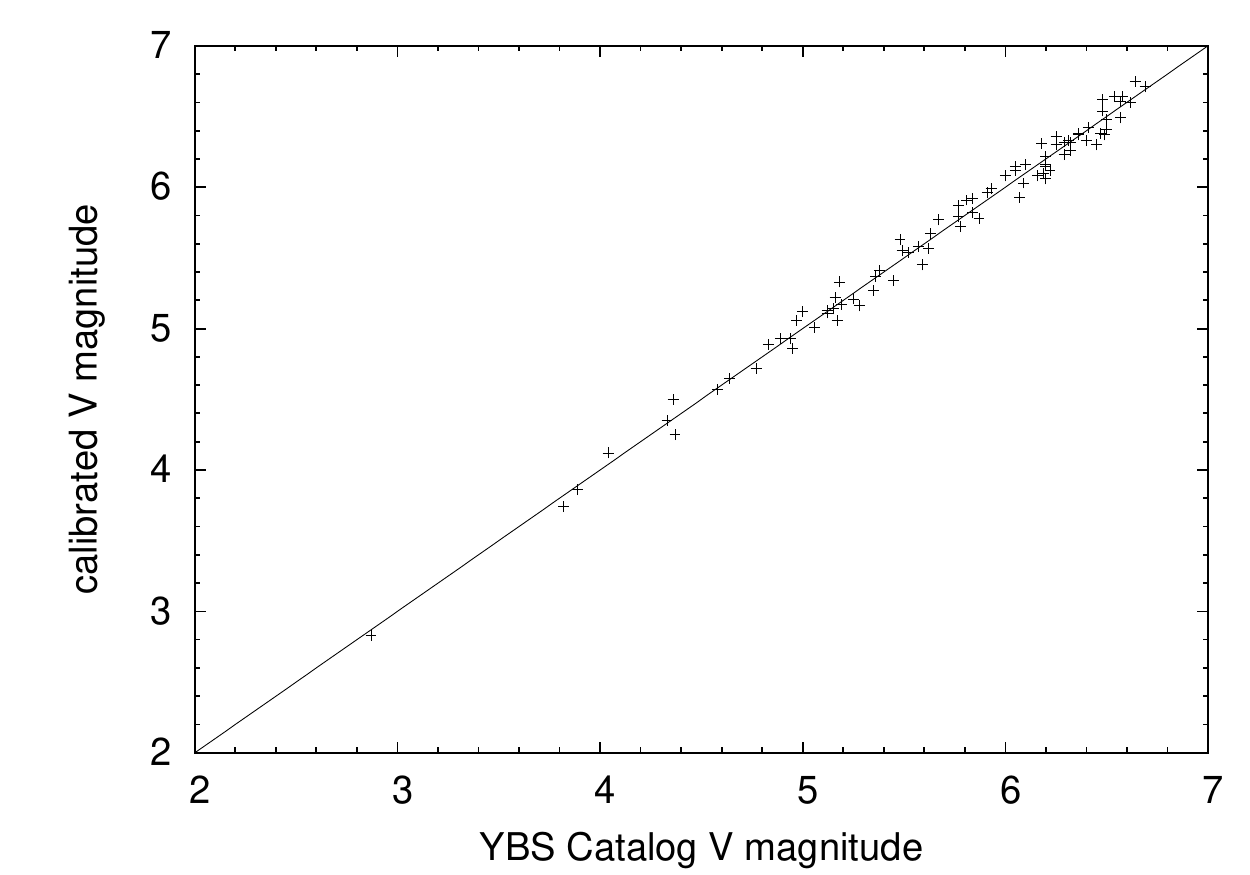}
\caption{Calibrated $V$ magnitudes (calculated the measured flux via to Eq.~(\ref{eq:fluxcal})) vs.\ YBS catalog magnitudes.}
\label{fig:fluxcal}
\end{figure}

\subsubsection{Standard star processing}
For each field and color channel, a list of several hundred reference stars (that included their equatorial coordinates, fluxes $S$ (in ADU) and local backgrounds $\beta$) was generated using Source Extractor and matched against the Yale Bright Star (YBS) Catalog~\citep{Hof1995}. Typically, 100\dots 120 matches were found. The reference star magnitude $V$ and the measured flux are related via
\begin{equation}
V = Q - 2.5\lg S + \epsilon(B-V)\:,
\label{eq:fluxcal}
\end{equation}
where $Q$ is the instrumental zero point (including effects of atmospheric extinction) and $(B-V)$ is the color index from the YBS Catalog. $Q$ and $\epsilon$ were then determined via a least squares fit. A typical plot of calibrated $V$ magnitudes (calculated according to Eq.~(\ref{eq:fluxcal})) vs.\ YBS catalog magnitudes is shown in Fig.~\ref{fig:fluxcal}.
As expected, the spectral correction parameter $\epsilon$ was close to $-1$ for the B channel frames and close to $0$ for the G channel, while values of approx.\ $+0.3$ were typical for the R channel. The ADU units were then converted to $S_{10}$(V) units commonly used for measuring surface brightness \citep{Spa1976}. One $S_{10}$(V) unit represents the sky brightness resulting from one $10^\text{th}$ magnitude star per square degree. Thus, 
\begin{equation}
1\text{~}S_{10}(\text{V}) = \kappa^2\, 10^{0.4(Q-10)}\text{~ADU}\:,
\end{equation}
where $\kappa$ is the image scale in degrees per pixel.

\subsubsection{Background subtraction}\label{sec:bgsubt}
According to the discussion at the start of this subsection, the total light intensity seen by a CCD pixel is 
\begin{equation}
I = S + I_\text{bad} + I_{good}\:,
\end{equation}
where $S$ is the flux due to point sources or resolved diffuse objects, $I_{bad} = I_\text{LP} + I_\text{A} + I_\text{ZL} + I_\text{sca}$ is the unwanted background contribution from terrestrial and solar system sources, and $I_\text{good} = I_\text{ISL} + I_\text{DGL} + I_\text{EBL}$ is the galactic and extragalactic background to be retained in the image. Since $I_\text{good}$ is known from the Pioneer 10/11 data \citep{Gor1998}, the spatial distribution of $I_\text{bad}$ can be modeled by fitting a surface to $(I-I_\text{good})$.

Unfortunately, the high-resolution background data by \cite{Gor1998} does not cover the entire sky; they excluded  data contaminated with scattered sunlight, taken within 70\degr\ of the Sun for Pioneer 10 and within 45\degr\ for Pioneer 11, respectively. \citet{Lei1998} provide some additional data points on a $10\degr\times 10\degr$ grid; the other missing data was interpolated using a bicubic spline function.
Some local median filtering was applied in order to remove small-scale artifacts in the data.

With the Pioneer data providing $I_\text{good}$, the ``bad'' background was estimated by applying a strong median filter to $(I-I_\text{good}) = (S + I_\text{bad})$. Subtracting $I_\text{bad}$ from $I$ then yields the desired sum of $S+I_\text{good}$. This approach also eliminates any residual gradient from the wavelength-dependent angular response of the camera system (cf.\ section~\ref{sec:preproc}).


%
%

\subsection{Final assembly and color processing}
The scaled and background-corrected images were resampled and co-added using SWarp. This software can be configured to generate a range of projections defined in the FITS/WCS standard \citep{Cal2002}, including Hammer-Aitoff, equidistant cylindrical (plate carr\'ee), and equidistant azimuthal. In addition, a number of target coordinate systems (equatorial, galactic, etc.) are available.

The resulting R, G and B fits images were converted to a single RGB TIFF file using the STIFF software \citep{Ber2008}. This program converts the 32-bit floating-point brightness values to 8-bit numbers by applying a non-linear transfer function to the luminance channel only. This way, bright objects do not appear ``burnt out'' as is often seen in images where the transfer function is applied directly to the R, G and B channels \citep{Lup2004}.

\section{Results}
The above procedure created a $360\degr\times 180\degr$ panorama image of the entire sky at a scale of 36\arcsec/pixel, equivalent to a size of $36000\times 18000$ pixels. With each pixel represented by a 32-bit (single precision) floating point number, each color channel is approx. 2.5~GByte in size.
The faintest visible stars are approx.\ 14~mag, while only stars brighter than $+0.7$~mag are saturated. (This saturation could have been avoided by reducing the shortest exposure time below 0.5~s, but this would have reduced the number of reference stars suitable for astrometric calibration.) Hence, the dynamic range is approx.\ 18 bits.

\begin{figure*}
{\centering\includegraphics[height=0.8\textheight]{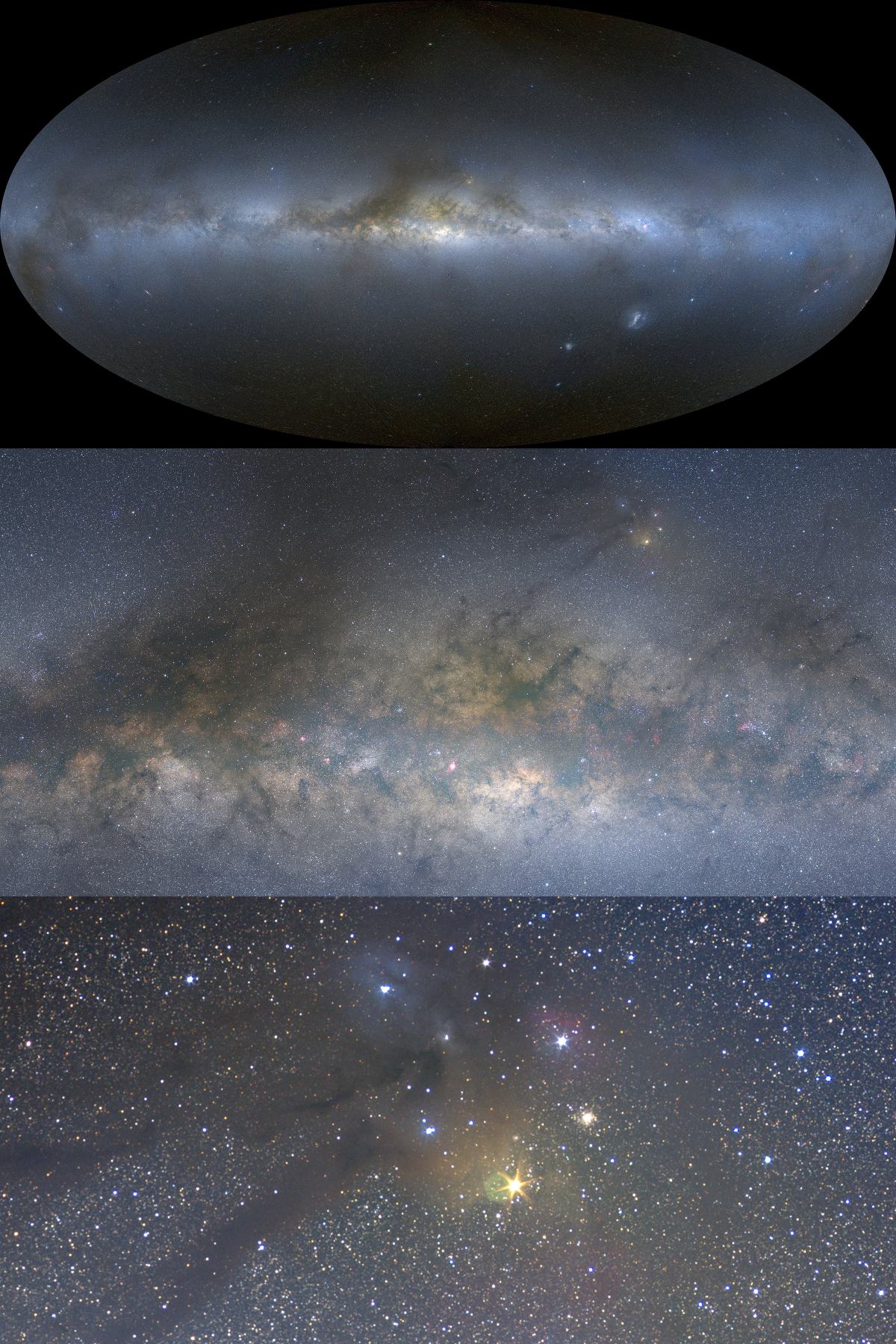}\\}
\caption{RGB mosaic, combined with photometric calibration and background subtraction. Top: entire sky in Hammer-Aitoff projection. Middle: enlarged view of the central Milky Way. Bottom: full-scale view of the Antares/$\rho$~Oph region.}
\label{fig:panorama}
\end{figure*}
A low-resolution color image of the panorama is shown in Fig.~\ref{fig:panorama}. A zoomable Mercator projection can be seen on the author's web site~\citep{mwpan2_web}. The image provides a detailed view of large-scale galactic structures, such as a halo of unresolved stars extending to $b\le \pm 25\degr$, and numerous dark clouds. Note the strong reddening of starlight in the Ophiuchus/Scorpius region of the Milky Way. In the earlier panorama \citep{Mel2005}, many of these details are missing due to the rather aggressive flat-fielding employed at the time.

Besides galactic stars and nebulae, and some brighter extragalactic objects (most notably, the Magellanic Clouds and the Andromeda Galaxy), the panorama image also shows a number of solar system objects. In most cases, the bright planets could be avoided by scheduling the exposures for times when no planet was near the field of view. No attempt was made to remove or avoid solar system objects invisible to the naked eye.

Additionally, several of the raw images taken from South Africa in April 2008 exhibit trails of geosynchronous satellites. At that time (2-3 weeks after the southern hemisphere fall equinox), sunlight glinting off reflective surfaces is directed towards the southern hemisphere, causing the satellite to brighten by several magnitudes. As the camera was tracking the stars, each satellite left a 1\degr\ trail on the 240~s exposures. Median-combining the raw frames eliminated most of the satellite trails; however, fields \#34 and 36 had to be re-taken as the blending of multiple trails caused them to be quite apparent even in the combined images.

\section{Applications}
First and foremost, the all-sky mosaic image is expected to be a valuable educational tool. Modern digital fulldome planetarium projection systems offer 12~bits of resolution per color channel, with 16~bits not too far down the road. With an 18-bit dynamic range, it will be possible to blend over from a view showing only the brightest stars (simulating the view in a light-polluted environment) to a full display of the faint star clouds and nebulae in our Milky Way. 
The resolution of $36\arcsec$/pixel ensures crisp star images that will be limited by the resolution of the projection equipment and the observer's eye, rather than the image data. As the size of the image (7.5~GByte for a 32-bit floating point representation) may exceed the capabilities of some personal computers, an OpenEXR version is in preparation. This file format uses 16-bit (half-precision) floating point numbers for storing the image data \citep{OpenEXR}, thus reducing the file size by a factor of two.

In addition to its use in education, the image may provide useful data on the Milky Way brightness in studies of the night sky background, such as \citet{Dur2007}.

\acknowledgments
The author is grateful to William Wren, Aret and Lampies Lambrechts, the staff of Big Bend National Park and the partners of Cederberg Astronomical Observatory, South Africa, for their hospitality and permission to image from their facilities.

\bibliography{AllSkyPanorama}

\begin{thebibliography}{28}
\expandafter\ifx\csname natexlab\endcsname\relax\def\natexlab#1{#1}\fi

\bibitem[{{Barnard} {et~al.}(1927){Barnard}, {Frost}, \& {Calvert}}]{Bar1927}
{Barnard}, E.~E., {Frost}, E.~B., \& {Calvert}, M.~R. 1927, {A photographic
  atlas of selected regions of the Milky way}, ed. E.~E. {Barnard}, E.~B.
  {Frost}, \& M.~R. {Calvert}

\bibitem[{Bertin(2008)}]{Ber2008}
Bertin, E. 2008, STIFF v1.12 User's guide,
  http://astromatic.iap.fr/software/stiff/

\bibitem[{{Bertin} \& {Arnouts}(1996)}]{Ber1996}
{Bertin}, E., \& {Arnouts}, S. 1996, \aaps, 117, 393

\bibitem[{{Bertin} {et~al.}(2002){Bertin}, {Mellier}, {Radovich}, {Missonnier},
  {Didelon}, \& {Morin}}]{Ber2002}
{Bertin}, E., {Mellier}, Y., {Radovich}, M., {Missonnier}, G., {Didelon}, P.,
  \& {Morin}, B. 2002, in Astronomical Society of the Pacific Conference
  Series, Vol. 281, Astronomical Data Analysis Software and Systems XI, ed.
  D.~A. {Bohlender}, D.~{Durand}, \& T.~H. {Handley}, 228

\bibitem[{Bortle(2001)}]{Bor2001}
Bortle, J.~E. 2001, Sky and Telescope, 101, 126

\bibitem[{{Calabretta} \& {Greisen}(2002)}]{Cal2002}
{Calabretta}, M.~R., \& {Greisen}, E.~W. 2002, \aap, 395, 1077

\bibitem[{{di Cicco}(1999)}]{diCicco1999}
{di Cicco}, D. 1999, \skytel, 98, 137

\bibitem[{Duriscoe {et~al.}(2007)Duriscoe, Luginbuhl, \& Moore}]{Dur2007}
Duriscoe, D.~M., Luginbuhl, C.~B., \& Moore, C.~A. 2007, \pasp, 119, 192

\bibitem[{{Eastman Kodak Company}(2006)}]{Kodak}
{Eastman Kodak Company}. 2006, Application note: KODAK KAI-11002 Image Sensor,
  Revision 1.0 MTD/PS-0938

\bibitem[{{Finkbeiner}(2003)}]{Fin2003}
{Finkbeiner}, D.~P. 2003, \apjs, 146, 407

\bibitem[{Gaustad {et~al.}(2001)Gaustad, McCullough, Rosing, \& {Van
  Buren}}]{SHASSA}
Gaustad, J.~E., McCullough, P.~R., Rosing, W., \& {Van Buren}, D. 2001, PASP,
  113, 1326

\bibitem[{{Gordon}(1997)}]{Gor1997}
{Gordon}, K.~D. 1997, PhD thesis, University of Toledo

\bibitem[{{Gordon} {et~al.}(1998){Gordon}, {Witt}, \& {Friedmann}}]{Gor1998}
{Gordon}, K.~D., {Witt}, A.~N., \& {Friedmann}, B.~C. 1998, \apj, 498, 522

\bibitem[{{Hoffleit} \& {Warren}(1995)}]{Hof1995}
{Hoffleit}, D., \& {Warren}, Jr., W.~H. 1995, {Yale Bright Star Catalogue, 5th
  Revised Ed.}, VizieR Online Data Catalog \#5050

\bibitem[{Kainz(2009)}]{OpenEXR}
Kainz, F. 2009, Technical Introduction to OpenEXR, Industrial Light \& Magic,
  \url{http://www.openexr.com}

\bibitem[{{Kizer Whitt} \& Mellinger(2001)}]{Kiz2001}
{Kizer Whitt}, K., \& Mellinger, A. 2001, Astronomy, 29, 58

\bibitem[{Leinert {et~al.}(1998)Leinert, Bowyer, Haikala, Hanner, Hauser,
  Levasseur-Regourd, Mann, Mattila, Reach, Schlosser, Staude1, Toller, Weiland,
  Weinberg, \& Witt}]{Lei1998}
Leinert, C., {et~al.} 1998, \aaps, 127, 1

\bibitem[{{Lupton} {et~al.}(2004){Lupton}, {Blanton}, {Fekete}, {Hogg},
  {O'Mullane}, {Szalay}, \& {Wherry}}]{Lup2004}
{Lupton}, R., {Blanton}, M.~R., {Fekete}, G., {Hogg}, D.~W., {O'Mullane}, W.,
  {Szalay}, A., \& {Wherry}, N. 2004, \pasp, 116, 133

\bibitem[{Mellinger(2008)}]{Mel2008}
Mellinger, A. 2008, in Handbook of Star Forming Regions, ed. B.~Reipurth, Vol.
  I: The Northern Sky (Astronomical Society of the Pacific)

\bibitem[{Mellinger(2009)}]{mwpan2_web}
---. 2009, Milky Way Panorama 2.0, \url{http://home.arcor.de/axel.mellinger/}

\bibitem[{Mellinger \& Hoffmann(2005)}]{Mel2005}
Mellinger, A., \& Hoffmann, S.~M. 2005, The New Atlas of the Stars (Firefly
  Books)

\bibitem[{{Minkowski} \& {Abell}(1963)}]{Min1963}
{Minkowski}, R.~L., \& {Abell}, G.~O. 1963, {The National Geographic
  Society-Palomar Observatory Sky Survey}, ed. K.~A. {Strand} (the University
  of Chicago Press), 481

\bibitem[{{Morrison} {et~al.}(2001){Morrison}, {R{\"o}ser}, {McLean},
  {Bucciarelli}, \& {Lasker}}]{Mor2001}
{Morrison}, J.~E., {R{\"o}ser}, S., {McLean}, B., {Bucciarelli}, B., \&
  {Lasker}, B. 2001, \aj, 121, 1752

\bibitem[{Richmond(2008)}]{Ric2008}
Richmond, M. 2008, Match -- a program for matching star lists,
  \url{http://spiff.rit.edu/match/}

\bibitem[{Shamir \& Nemiroff(2005)}]{Sha2005}
Shamir, L., \& Nemiroff, R.~J. 2005, \pasp, 117, 972

\bibitem[{{Sparrow} \& {Weinberg}(1976)}]{Spa1976}
{Sparrow}, J.~G., \& {Weinberg}, J.~L. 1976, in Lecture Notes in Physics,
  Vol.~48, Interplanetary Dust and Zodiacal Light, ed. H.~{Elsaesser} \&
  H.~{Fechtig} (Springer), 41--44

\bibitem[{{Valdes} {et~al.}(1995){Valdes}, {Campusano}, {Velasquez}, \&
  {Stetson}}]{Val1995}
{Valdes}, F.~G., {Campusano}, L.~E., {Velasquez}, J.~D., \& {Stetson}, P.~B.
  1995, \pasp, 107, 1119

\bibitem[{{York} {et~al.}(2000){York}, {Adelman}, {Anderson}, {Anderson},
  {Annis}, {Bahcall}, {Bakken}, {Barkhouser}, {Bastian}, {Berman}, {Boroski},
  {Bracker}, {Briegel}, {Briggs}, {Brinkmann}, {Brunner}, {Burles}, {Carey},
  {Carr}, {Castander}, {Chen}, {Colestock}, {Connolly}, {Crocker}, {Csabai},
  {Czarapata}, {Davis}, {Doi}, {Dombeck}, {Eisenstein}, {Ellman}, {Elms},
  {Evans}, {Fan}, {Federwitz}, {Fiscelli}, {Friedman}, {Frieman}, {Fukugita},
  {Gillespie}, {Gunn}, {Gurbani}, {de Haas}, {Haldeman}, {Harris}, {Hayes},
  {Heckman}, {Hennessy}, {Hindsley}, {Holm}, {Holmgren}, {Huang}, {Hull},
  {Husby}, {Ichikawa}, {Ichikawa}, {Ivezi{\'c}}, {Kent}, {Kim}, {Kinney},
  {Klaene}, {Kleinman}, {Kleinman}, {Knapp}, {Korienek}, {Kron}, {Kunszt},
  {Lamb}, {Lee}, {Leger}, {Limmongkol}, {Lindenmeyer}, {Long}, {Loomis},
  {Loveday}, {Lucinio}, {Lupton}, {MacKinnon}, {Mannery}, {Mantsch}, {Margon},
  {McGehee}, {McKay}, {Meiksin}, {Merelli}, {Monet}, {Munn}, {Narayanan},
  {Nash}, {Neilsen}, {Neswold}, {Newberg}, {Nichol}, {Nicinski}, {Nonino},
  {Okada}, {Okamura}, {Ostriker}, {Owen}, {Pauls}, {Peoples}, {Peterson},
  {Petravick}, {Pier}, {Pope}, {Pordes}, {Prosapio}, {Rechenmacher}, {Quinn},
  {Richards}, {Richmond}, {Rivetta}, {Rockosi}, {Ruthmansdorfer}, {Sandford},
  {Schlegel}, {Schneider}, {Sekiguchi}, {Sergey}, {Shimasaku}, {Siegmund},
  {Smee}, {Smith}, {Snedden}, {Stone}, {Stoughton}, {Strauss}, {Stubbs},
  {SubbaRao}, {Szalay}, {Szapudi}, {Szokoly}, {Thakar}, {Tremonti}, {Tucker},
  {Uomoto}, {Vanden Berk}, {Vogeley}, {Waddell}, {Wang}, {Watanabe},
  {Weinberg}, {Yanny}, \& {Yasuda}}]{York2000}
{York}, D.~G., {et~al.} 2000, \aj, 120, 1579

\end{thebibliography}

\clearpage

\begin{deluxetable}{lp{7cm}rrr}
\tablecaption{Observation sites\label{tab:sites}}
\tablehead{\colhead{Site abbrev.} & \colhead{Site} & \colhead{latitude} & \colhead{longitude} & elevation (m)}
\tablewidth{0pt}
\startdata
CEDB & Cederberg Observatory, South Africa      & $32\degr 29\arcmin 58\arcsec$ S& $19\degr 15\arcmin 10\arcsec$ E & 865\\
KOOR & Koornlandskloof Guest Farm,\newline South Africa & $32\degr 13\arcmin 34\arcsec$ S& $20\degr 20\arcmin 48\arcsec$ E & 1273\\
BBE1 & Big Bend National Park, Texas, USA\newline Rio Grande Village & $29\degr 10\arcmin 45\arcsec$ N& $102\degr 57\arcmin 20\arcsec$ W & 556 \\
BBE2 & Big Bend National Park, Texas, USA\newline Grapevine Hills & $29\degr 22\arcmin 40\arcsec$ N& $103\degr 13\arcmin 16\arcsec$ W& 1029 \\
BBE3 & Big Bend National Park, Texas, USA\newline Terlingua Abajo & $29\degr 11\arcmin 57\arcsec$ N& $103\degr 36\arcmin 14\arcsec$ W& 677\\
FTDV & near Fort Davis, Texas, USA & $30\degr 36\arcmin 54\arcsec$ N& $103\degr 58\arcmin 38\arcsec$ W & 1580 \\
FRMI & Filion Rd.\ near Bay Port, MI, USA & $43\degr 53\arcmin 12\arcsec$ N & $83\degr 19\arcmin 53\arcsec$ W & 177\\
MCLL & McCollum Lake, MI, USA & $44\degr 46\arcmin 13\arcsec$ N & $83\degr 54\arcmin 05\arcsec$ W & 285\\
HNFO & Huron National Forest, MI, USA & $44\degr 39\arcmin 09\arcsec$ N & $83\degr 51\arcmin 47\arcsec$ W & 290
\enddata
\end{deluxetable}

\begin{deluxetable}{ll@{\hspace{8ex}}rr@{\hspace{8ex}}cc}
\tablecaption{Field list. A few images were affected by the presence of bright planets or satellite trails and were re-taken. In these cases, data from both sets of data was used for correcting the defects.}
\tablecolumns{6}
\tablewidth{0pt}
\tablehead{\colhead{Field} & \colhead{Site} & \multicolumn{2}{c}{image center} & \multicolumn{2}{c}{start of first exposure (UTC)} \\
 & & RA & Dec & date & time}
\startdata
00 &	BBE2 &	$ 0^\text{h} 00^\text{m}$ & $+90\degr$ &	2008-12-27 &	05:47:01\\
01 &	BBE1 &	$ 0^\text{h} 00^\text{m}$ & $+62\degr$ &	2008-12-26 &	03:38:41\\
02 &	BBE1 &	$ 2^\text{h} 24^\text{m}$ & $+62\degr$ &	2008-12-26 &	05:15:59\\
03 &	BBE2 &	$ 4^\text{h} 48^\text{m}$ & $+62\degr$ &	2008-12-26 &	03:56:00\\
04 &	FTDV &	$ 7^\text{h} 12^\text{m}$ & $+62\degr$ &	2008-12-30 &	05:54:20\\
05 &	BBE1 &	$ 9^\text{h} 36^\text{m}$ & $+62\degr$ &	2008-12-25 &	11:03:10\\
06 &	BBE2 &	$12^\text{h} 00^\text{m}$ & $+62\degr$ &	2008-12-27 &	10:58:39\\
07 &	BBE3 &	$14^\text{h} 24^\text{m}$ & $+62\degr$ &	2008-12-28 &	09:28:14\\
08 &	MCLL &	$16^\text{h} 48^\text{m}$ & $+62\degr$ &	2009-05-25 &	03:05:09\\
09 &	HNFO &	$19^\text{h} 12^\text{m}$ & $+62\degr$ &	2009-07-21 &	03:02:54\\
10 &	BBE2 &	$21^\text{h} 36^\text{m}$ & $+62\degr$ &	2008-12-27 &	02:14:30\\
11 &	BBE1 &	$00^\text{h} 00^\text{m}$ & $+32\degr$ &	2008-12-23 &	01:56:22\\
12 &	BBE1 &	$01^\text{h} 36^\text{m}$ & $+32\degr$ &	2008-12-23 &	03:42:07\\
13 &	BBE1 &	$03^\text{h} 12^\text{m}$ & $+32\degr$ &	2008-12-23 &	05:28:04\\
14 &	BBE1 &	$04^\text{h} 48^\text{m}$ & $+32\degr$ &	2008-12-24 &	03:56:22\\
15 &	BBE1 &	$06^\text{h} 24^\text{m}$ & $+32\degr$ &	2008-12-23 &	07:11:20\\
16 &	BBE1 &	$08^\text{h} 00^\text{m}$ & $+32\degr$ &	2008-12-23 &	08:53:32\\
17 &	BBE1 &	$09^\text{h} 36^\text{m}$ & $+32\degr$ &	2008-12-24 &	07:24:30\\
18 &	BBE1 &	$11^\text{h} 12^\text{m}$ & $+32\degr$ &	2008-12-24 &	09:04:45\\
19 &	BBE1 &	$12^\text{h} 48^\text{m}$ & $+32\degr$ &	2008-12-24 &	10:36:04\\
20 &	BBE3 &	$14^\text{h} 24^\text{m}$ & $+32\degr$ &	2008-12-28 &	11:04:58\\
21 &	MCLL &	$16^\text{h} 00^\text{m}$ & $+32\degr$ &	2009-05-25 &	04:48:28\\
22 &	MCLL &	$17^\text{h} 36^\text{m}$ & $+32\degr$ &	2009-05-25 &	06:14:52\\
23 &	HNFO &	$19^\text{h} 12^\text{m}$ & $+32\degr$ &	2009-07-21 &	04:45:24\\
24 &	HNFO &	$20^\text{h} 48^\text{m}$ & $+32\degr$ &	2009-08-16 &	03:03:46\\
25 &	BBE3 &	$22^\text{h} 24^\text{m}$ & $+32\degr$ &	2008-12-28 &	01:36:06\\
26 &	KOOR &	$ 0^\text{h} 00^\text{m}$ & $ 00\degr$ &	2007-10-10 &	20:12:49\\
27 &	KOOR &	$ 1^\text{h} 20^\text{m}$ & $ 00\degr$ &	2007-10-12 &	20:21:02\\
28 &	KOOR &	$ 2^\text{h} 40^\text{m}$ & $ 00\degr$ &	2007-10-09 &	23:55:27\\
29 &	KOOR &	$ 4^\text{h} 00^\text{m}$ & $ 00\degr$ &	2007-10-10 &	01:56:45\\
30 &	KOOR &	$ 5^\text{h} 20^\text{m}$ & $ 00\degr$ &	2007-10-13 &	01:40:04\\
31 &	KOOR &	$ 6^\text{h} 40^\text{m}$ & $ 00\degr$ &	2008-04-07 &	17:48:53\\
32 &	KOOR &	$ 8^\text{h} 00^\text{m}$ & $ 00\degr$ &	2008-04-08 &	17:50:36\\
33 &	KOOR &	$ 9^\text{h} 20^\text{m}$ & $ 00\degr$ &	2008-04-07 &	19:17:06\\
34 &	KOOR &	$10^\text{h} 40^\text{m}$ & $ 00\degr$ &	2008-04-05 &	19:43:51\\
34b&	FRMI &	$10^\text{h} 40^\text{m}$ & $ 00\degr$ &	2009-04-25 &	02:47:36\\
35 &	KOOR &	$12^\text{h} 00^\text{m}$ & $ 00\degr$ &	2008-04-05 &	20:46:28\\
36 &	KOOR &	$13^\text{h} 20^\text{m}$ & $ 00\degr$ &	2008-04-05 &	22:10:40\\
36b&	FTDV &	$13^\text{h} 20^\text{m}$ & $ 00\degr$ &	2008-12-30 &	11:18:56\\
37 &	KOOR &	$14^\text{h} 60^\text{m}$ & $ 00\degr$ &	2008-04-07 &	21:13:17\\
38 &	KOOR &	$16^\text{h} 00^\text{m}$ & $ 00\degr$ &	2008-04-07 &	22:56:53\\
39 &	KOOR &	$17^\text{h} 20^\text{m}$ & $ 00\degr$ &	2008-04-08 &	00:34:04\\
40 &	KOOR &	$18^\text{h} 40^\text{m}$ & $ 00\degr$ &	2007-10-12 &	18:19:46\\
41 &	KOOR &	$20^\text{h} 00^\text{m}$ & $ 00\degr$ &	2007-10-09 &	18:20:18\\
42 &	KOOR &	$21^\text{h} 20^\text{m}$ & $ 00\degr$ &	2007-10-09 &	20:02:07\\
43 &	KOOR &	$22^\text{h} 40^\text{m}$ & $ 00\degr$ &	2007-10-09 &	21:40:59\\
44 &	KROM &	$00^\text{h} 00^\text{m}$ & $-32\degr$ &	2007-10-04 &	20:27:47\\
45 &	KROM &	$01^\text{h} 36^\text{m}$ & $-32\degr$ &	2007-10-04 &	22:22:48\\
46 &	KROM &	$03^\text{h} 12^\text{m}$ & $-32\degr$ &	2007-10-05 &	00:05:33\\
47 &	KOOR &	$04^\text{h} 48^\text{m}$ & $-32\degr$ &	2007-10-12 &	22:21:22\\
48 &	KOOR &	$06^\text{h} 24^\text{m}$ & $-32\degr$ &	2007-10-12 &	23:59:58\\
49 &	CEDB &	$08^\text{h} 00^\text{m}$ & $-32\degr$ &	2008-03-29 &	19:14:05\\
49 &	KOOR &	$08^\text{h} 00^\text{m}$ & $-32\degr$ &	2008-04-08 &	19:37:29\\
50 &	CEDB &	$09^\text{h} 36^\text{m}$ & $-32\degr$ &	2008-03-29 &	21:23:53\\
51 &	CEDB &	$11^\text{h} 24^\text{m}$ & $-32\degr$ &	2008-03-30 &	19:56:46\\
51 &	KOOR &	$11^\text{h} 24^\text{m}$ & $-32\degr$ &	2008-04-03 &	20:18:02\\
52 &	KOOR &	$12^\text{h} 48^\text{m}$ & $-32\degr$ &	2008-04-08 &	22:58:17\\
53 &	KOOR &	$14^\text{h} 24^\text{m}$ & $-32\degr$ &	2008-04-09 &	00:46:19\\
54 &	KOOR &	$16^\text{h} 00^\text{m}$ & $-32\degr$ &	2008-04-05 &	23:45:24\\
55 &	KOOR &	$17^\text{h} 36^\text{m}$ & $-32\degr$ &	2008-04-06 &	01:34:05\\
55b &	KOOR &	$17^\text{h} 36^\text{m}$ & $-32\degr$ &	2008-04-08 &	02:38:10\\
56 &	CEDB &	$19^\text{h} 12^\text{m}$ & $-32\degr$ &	2007-10-05 &	18:06:31\\
57 &	CEDB &	$20^\text{h} 48^\text{m}$ & $-32\degr$ &	2007-10-05 &	20:02:16\\
58 &	KROM &	$22^\text{h} 24^\text{m}$ & $-32\degr$ &	2007-10-04 &	18:39:01\\
59 &	CEDB &	$00^\text{h} 00^\text{m}$ & $-62\degr$ &	2007-10-05 &	22:01:34\\
60 &	CEDB &	$02^\text{h} 24^\text{m}$ & $-62\degr$ &	2007-10-05 &	23:45:45\\
61 &	KOOR &	$04^\text{h} 48^\text{m}$ & $-62\degr$ &	2008-04-05 &	17:58:42\\
62 &	KOOR &	$07^\text{h} 12^\text{m}$ & $-62\degr$ &	2008-04-03 &	18:01:53\\
63 &	KOOR &	$09^\text{h} 36^\text{m}$ & $-62\degr$ &	2008-04-08 &	20:55:47\\
64 &	KOOR &	$12^\text{h} 00^\text{m}$ & $-62\degr$ &	2008-04-01 &	23:47:05\\
65 &	KOOR &	$14^\text{h} 24^\text{m}$ & $-62\degr$ &	2008-04-03 &	23:26:34\\
66 &	KOOR &	$16^\text{h} 48^\text{m}$ & $-62\degr$ &	2008-04-04 &	01:03:40\\
67 &	KOOR &	$19^\text{h} 12^\text{m}$ & $-62\degr$ &	2007-10-08 &	18:13:48\\
68 &	KOOR &	$21^\text{h} 36^\text{m}$ & $-62\degr$ &	2007-10-08 &	19:52:48\\
69 &	KOOR &	$00^\text{h} 00^\text{m}$ & $-90\degr$ &	2007-10-08 &	21:41:51
\enddata
\end{deluxetable}

\end{document}